\documentstyle[fullpage,fleqn]{article}
\font\fonta=cmr12 scaled\magstep2

\font\fontc=cmr12 scaled\magstep1

\def\ni {\noindent}
\begin{document}
{
\begin{flushright}{TMU-NT940101\\January 1994}
\end{flushright}
}
\vspace {0.3cm}
\large
\baselineskip=1cm
\begin{center}{\fonta Meson Structure in Deep Inelastic Scattering} \\
\vspace {1.5cm}
{\fontc Takayuki Shigetani,  Katsuhiko Suzuki
\footnote{\large{e-mail address :
ksuzuki@atlas.phys.metro-u.ac.jp}}
and  Hiroshi Toki
\footnote{{\large also RIKEN, Wako, Saitama, 351, Japan}}\\
        {\em Department of Physics, Tokyo Metropolitan University}\\
        {\em Hachiohji,Tokyo 192-03, Japan}}

\vspace {3cm}
{\bf Abstract}
\end{center}
\vspace {0.5cm}

\baselineskip=0.8cm
\noindent
We study the deep inelastic structure functions of
mesons within the Nambu and Jona-Lasinio model.  We calculate the
valence quark distributions in $\pi $, $K $, and $\rho $ mesons
at the low
energy model scale, which are evoluted to the experimental momentum
scale in terms of the Altarelli-Parisi equation.
The resulting distribution functions show reasonable agreements with
experiment.  We also discuss the semi-inclusive lepton nucleon
scattering process with a slow nucleon in coincidence
in the final state, which reveals the off-shell structure of the pion.
\newpage
\baselineskip=0.8cm
\noindent
{\fontc {\bf 1 Introduction}}

The deep inelastic scattering (DIS) of hadrons is one of the most
powerful tool to investigate their internal quark structure.
In the last decade, the
nucleon structure function has been extensively studied, and provided
us with detailed information of the nucleon structure\cite{Roberts}.
The observed
scaling violations of the structure functions are consistent with the
perturbative QCD predictions, though at the present QCD can not predict
the structure function itself.

On the other hand, it is of great interest to clarify a connection
between DIS information and the low energy quark models\cite{Jaffe}.
At the experimental large momentum scale, the virtual photon sees the
hadron as a complicated object, which consists of valence quarks,
sea quarks, and gluons.  As the momentum
scale becomes smaller, sea quarks and
gluons are absorbed into valence quarks, and their degrees of freedom
are substituted by the 'constituent quarks', whose dynamics is subject
to the low energy QCD.
If we relate the DIS data with the quark models, we can learn how the
non-perturbative aspects of QCD, e.g. confinement and chiral symmetry
breaking, reflect the behavior of the quark distributions at the DIS
scale.
Recently, such theoretical studies for the nucleon structure function
are made in terms
of several effective quark models, which are supposed to work at the
low energy scale\cite{Miller,Thomas,Models}.
Those works are based on the assumption that
the structure functions at the low energy model scale $Q^2=Q_0^2$
are obtained by calculating the twist-2 matrix elements within the
effective models.  In the formalism of the operator product
expansion(OPE), the $n$-th moment of the structure function
$F_2$ at the scale $Q^2$ is expanded as,
%
%
\begin{eqnarray}
\int {dx}x^{n-2}F_{2}(x,Q^{2}) = \sum_{\tau} C_{\tau}^{n}
(Q^{2},Q^{2}_{0})
<{\cal O}_{\tau}^{n}(Q^{2}_{0})>
\end{eqnarray}
\noindent
where $C_{\tau}^{n}(Q^{2},Q^{2}_{0})$ is the Wilson coefficients
calculated by the perturbative QCD, and
$<{\cal O}_{\tau}^{n}(Q^{2}_{0})>$ is the expectation value of the
local operators which are evaluated at the arbitrary scale $Q^{2}_{0}$.
In the Bjorken
limit ($Q^2\rightarrow \infty$), only the twist-2 ($\tau=2$) term
survives and the higher twist terms become negligible
($\sim O (1/Q^{2})$).
Hence, once we calculate the twist-2 operators within the quark model
at the low momentum scale $Q^{2}_{0}$ where
the phenomenological model makes sense, we can get
the structure function at the experimental scale $Q^{2}$ using the
QCD evolution equation.
Thus, the comparison with experiment can be made.

	Although the available experimental data of meson structure
function is much fewer than that of nucleon,
the quark distributions
in mesons are also important to study the quark structure of hadrons.
Comparing with the nucleon case, one may extract more directly the
information of the quark-quark interaction
from the structure function,
since we can avoid solving the complicated three body problem as in
the nucleon case.  In this paper, we shall calculate the quark
distributions in mesons within the Nambu and Jona-Lasinio (NJL)
model\cite{NJL}.
In this model, the gluon degrees of freedom are assumed
to be frozen into a chiral invariant effective 4-point interaction in
the low energy region\cite{Effective}.  The NJL model demonstrates the
spontaneous breakdown of chiral symmetry and the emergence of the
Goldstone bosons.  The generalized $SU(3)_f $ NJL model reproduces
the meson properties remarkably well, in spite of the lack of
confinement\cite{Vogl,Review}.
This model is also applied to the chiral phase transition at finite
temperature and density\cite{Review}.  All these results indicate
that the NJL model possesses the essential features of QCD.
Note that the NJL model satisfies the Lorentz invariance and certain
kinematics.  This is an advantage of our use of the NJL model,
because the quark model which does not satisfy the translational
invariance, e.g. the MIT bag model, gives an incorrect behavior
of the quark distributions at $x=x_B=1 $\cite{Miller,Thomas}.
This difficulty, so called the support
problem, may be removed by the momentum projection.  However, some
ambiguities exist in the projection methods\cite{PT}.

	In the recent work of present authors\cite{Shige}, the pion and
kaon structure functions were calculated in the NJL model, and our
results turned out to be consistent with experiment.
In that work, however, we used
an {\it artificial} momentum cutoff to reproduce the experimentally
known behavior around $x\sim 1$.  Here, we newly fix the momentum
cutoff procedure to reproduce the meson properties such as masses
and decay constants, and avoid the ambiguity.

	We also study the semi-inclusive lepton nucleon scattering as a
possible way to measure the quark distributions in mesons.
Experimentally, the pion structure function has been extracted from
the Drell-Yan process\cite{DrellYan,ratio}.
On the other hand, the semi-inclusive processes, e.g.
$p+\ell\to n+\ell'+X$, are expected to allow a new determination of the
quark distributions in the pion\cite{Semi1,Semi2,Semi3}.
In such processes, the virtual photon
probes the virtual meson clouds around the nucleon, especially pion.
Hence, this experiment will determine the {\it off-shell}
pion structure function.
We can make predictions for such semi-inclusive scattering in terms of
the calculated pion structure function.

	This paper is arranged as follows.  In sec. 2, the Nambu and
Jona-Lasinio model is introduced, and its model parameters are fixed to
reproduce the meson properties.  Sec. 3 is devoted to the calculation
of the hadronic tensor of mesons.  The structure functions of the
pseudoscalar channel as well as the vector channel are given in the
NJL model, where twist-2 contributions are calculated in the Bjorken
limit.  These quark distributions are evoluted to the experimental high
$Q^2 $ scale with the help of the Altarelli-Parisi equation in sec. 4.
We compare them with data from the Drell-Yan experiment.
The semi-inclusive processes are formulated in
sec. 5.  The final section is devoted to summary and discussions.
\vspace {3cm}

\noindent
{\fontc {\bf 2 The Nambu and Jona-Lasinio model}}

The Nambu and Jona-Lasinio model\cite{NJL} has attracted considerable
interests as a low energy effective theory of QCD.  The NJL model
incorporates the chiral symmetry, which is one of the most important
aspects of the low energy QCD, and provides the simple picture of its
spontaneous and explicit breakdown.  Recently, this model
was extensively studied in several subjects of hadron physics, and gave
successful results\cite{Vogl,Review}.  The generalized $SU(3)_{f}$
NJL model lagrangian is given by,
%
%
\vglue 0.1cm
\begin{eqnarray}
{\cal L}_{NJL}&=&{\cal L}_{o} + {\cal L}_{4}\nonumber\\
{\cal L}_{o}  &=& \bar{\psi}(i\gamma^{\mu}\partial_{\mu} - m)
\psi\nonumber\\
{\cal L}_{4}  &=& G_{S}[(\bar{\psi}t_{a}\psi)^2
                 +(\bar{\psi}t_{a}i\gamma^{5}\psi)^2]\nonumber\\
              & &-G_{V}[(\bar{\psi}t_{a}\gamma_{\mu}\psi)^2
                 +(\bar{\psi}t_{a}\gamma_{\mu}\gamma^{5}\psi)^2] \; .
\label{lag}
\end{eqnarray}
\vglue 0.3cm
\noindent
Here, $\psi$ denotes the quark field with the current mass $m$.
$t_i$ are the $SU(3)$ flavor matrices, normalized as
$tr(t_{i}t_{j})=\delta_{ij}/2$,
and $G_S$, $G_V$ the coupling constants.

Using (\ref{lag}), one gets the Dyson equation for the quark
propagator in the Hartree approximation (Fig.1 (a));
%
%
\begin{eqnarray}
M=m-G_S<\bar \psi \psi > \; ,
\label{gap}
\end{eqnarray}
\noindent
where $M$ is the dynamically generated constituent quark mass.  The
quark condensate $<{\bar \psi}\psi>$ is expressed as,
%
%
\begin{eqnarray}
<\bar \psi \psi >=-i4N_c\int {{{d^4p} \over {(2\pi )^4}}}{M \over
{p^2-M^2}} \; ,
\end{eqnarray}
\vglue 0.5cm
\noindent
where $N_{c}$ is the number of colors.
If the coupling constant $G_{S}$ is larger than a critical value,
the quark condensate and the constituent mass have non-zero values,
which means spontaneous breakdown of the chiral symmetry.

The meson properties are obtained by solving the Bethe-Salpeter (BS)
equation, illustrated in Fig.1 (b).  The inhomogeneous BS equation for
$q$-$\bar q$ T-matrix is written symbolically as,
%
%
\begin{eqnarray}
{\cal T \, = \, K \, + \, KJT } \; ,
\label{T-matrix}
\end{eqnarray}
\noindent
where ${\cal T}$ is the T-matrix, ${\cal K}$ is the interaction kernel
obtained from (\ref{lag}), and ${\cal J}$ the loop integral of a given
channel.  ${\cal K}$ is decomposed into the various products of the
gamma, flavor and color matrices.(See ref.\cite{Vogl} for detail.)
The loop integral ${\cal J}$ is defined by,
%
%
\vglue 0.2cm
\begin{eqnarray}
{\cal J}_{ij}(q,M_1,M_2)=i\int {{{d^4k} \over {(2\pi )^4}}}
tr[S(k+\frac{1}{2}q,M_1)\Gamma _iS(k- \frac{1}{2}q,M_2)\Gamma _j]\; ,
\label{loop}
\end{eqnarray}
\vglue 0.5cm
\noindent
where \(S_{F}(k,M)\) is the quark propagator.
%
%
\begin{eqnarray*}
S_{F}(k,M)=\frac{1}{{\not k}-M}
\end{eqnarray*}
\vglue 0.3cm
\noindent
The quark propagator contains the constituent mass obtained by solving
the gap equation (\ref{gap}).
The meson mass is determined as a pole of T-matrix at
$q^{2} = m_{meson}^2$.  The quark-antiquark T-matrix
(\ref{T-matrix}) is
rewritten as,
%
%
\begin{eqnarray}
{\cal T}(q) & = & {\cal K \, + \, KJT }\nonumber\\
& = & \frac {{\cal K}} {1 - {\cal JK}}\nonumber\\
&\sim & g_{m qq} \Gamma_{i} \, \frac {1} {\, q^2 - m_{meson}^2}
\, \Gamma_{j} g_{m qq}  \hspace {1cm} at \hspace {1cm}
 q^2\sim m_{meson}^2 \; \; .
\label{T-matrix-ex}
\end{eqnarray}
\vglue 0.5cm
\noindent
Here, $g_{m qq}$ is the meson-quark-quark coupling constant.
The integrals in (\ref{gap}) and (\ref{loop}) diverge due to the
non-renormalizablity of the NJL model.  Thus, we introduce
the Fermi-distribution type momentum cutoff function in these
integrals;
\begin{eqnarray}
\int {{d^4k}} \rightarrow
  i \int {{d^4k_E}}\frac {1}{1+exp[(k_{E}^{2} - \Lambda^2)/a]} \; .
\label{cutoff}
\end{eqnarray}
\vglue 0.2cm
\noindent
Here, $k_{E}^{2}$ is the Euclidean four momentum square, and $\Lambda$
is identified with the typical scale of the chiral
symmetry breaking $\sim 1GeV$.  We use $a\sim0.1GeV^2$ to reproduce
the meson properties.
This form of the regularization function is consistent with the
usual Euclidean sharp cutoff scheme\cite{Vogl}.  One may understand
the physical implication of the cutoff procedure as an approximate
realization of "Asymptotic Freedom" in the NJL model, i.e.
the interaction between two quarks with the relative
momentum larger than $\Lambda$ is turned off, and two particles are
free in such a high momentum region.
Hence, the structure functions obtained in the NJL model exhibit the
exact Bjorken scaling in the deep inelastic limit\cite{BJM}.

Solving the BS equation (\ref{T-matrix}), we can obtain the meson
masses, and the decay constants.  We take the current mass of
the up and down quarks $m_{u,d} = 5.5MeV$ as a semi-empirical value.
The remaining parameters are fixed by the pseudoscalar meson properties.
The parameters and the resulting physical properties are tabulated in
Table 1.  Using this model, we shall describe the deep inelastic
structure of mesons.

\vglue 3cm
\noindent
{\fontc {\bf 3 Calculation of the Hadronic Tensor}}

We evaluate the structure functions of mesons in the NJL model,
as done in ref.\cite{Shige}.  The calculated distribution gives a
boundary condition for the QCD perturbation at the model scale.
The hadronic tensor
\(W_{\mu\nu}\) is written by the structure functions
\(F_1 (x)\) and \(F_2 (x)\) in the scaling limit;
%
%
\begin{eqnarray}
W_{\mu\nu}=
&-&(g_{\mu\nu}-\frac{q_{\mu}q_{\nu}}{q^2})F_{1}(x)\nonumber\\
&+&\frac{1}{m_{ps}\nu}(p_{\mu}-\frac{p\cdot q}{q^2}q_{\mu})
                     (p_{\nu}-\frac{p\cdot q}{q^2}q_{\nu})F_{2}(x)
\end{eqnarray}
\vglue 0.5cm
\noindent
where
%
%
\begin{eqnarray*}
F_{2}(x)=x\sum_{i}e_{i}^{2}[q_{i}(x)+\bar{q}_{i}(x)]\; ,\;\;
F_{1}(x)=\frac{1}{2x}F_{2}(x)
\end{eqnarray*}
\vglue 0.5cm
\noindent
\(q_{i}(x)\) and \(\bar{q}_{i}(x)\) are the momentum distributions of
i-flavor quark and antiquark.
The hadronic tensor is related to the forward scattering amplitude
\(T_{\mu\nu}\) through the optical theorem\cite{Roberts}.
%
%
\begin{eqnarray}
W_{\mu\nu}=\frac{1}{2\pi}ImT_{\mu\nu}\; .
\label{dispersion}
\end{eqnarray}
\vglue 0.3cm
\noindent
Thus, we calculate \(T_{\mu\nu}\) in the NJL model to get
the structure functions.

We first consider the pseudoscalar meson case.  We compute the forward
scattering amplitude in the impulse approximation, which is
illustrated in Fig.2 ("handbag diagrams").  Here, $p$ is the momentum
of a target meson, and $q$ the momentum delivered by the virtual
photon.  We expand the matrix elements of Fig.2 as a series of
$1/Q^2$.  The leading $O((1/Q^{2})^{0})$ term of the "handbag diagram"
coincides with the twist-2 contribution of the
OPE in the Bjorken limit\cite{Rujula}.  Thus, this diagram is enough
for our purpose.  We note that, though leading terms do not
depend on $Q^2$ explicitly, they receive the logarithmic QCD radiative
corrections, which are incorporated by the Altarelli-Parisi equation.
The matrix elements of Fig.2 are given by,
%
%
\vglue 0.3cm
\begin{eqnarray}
T_{\mu\nu}=i\int \frac{d^{4}k}{(2\pi)^4}
Tr[\gamma_{\mu}Q\frac{1}{{\not k}}\gamma_{\nu}QT_{-}]+(T_{+} \; term),
\end{eqnarray}
\noindent
where
%
\begin{eqnarray}
T_{-}=S_{F}(k-q,M_1)g_{p qq}\tau_{+}i\gamma_{5}S_{F}(k-q-p,M_2)
g_{p qq}\tau_{-}i\gamma_{5}S_{F}(k-q,M_1)
\label{Tpion}
\end{eqnarray}
\vglue 0.5cm
\noindent
\(T_{-}\) represents the contribution with an antiquark being a
spectator.  \(T_{+}\) also has a similar expression, where a quark is a
spectator.
Here, \(g_{p qq}\) is the coupling constant among two quarks and the
pseudoscalar meson obtained from (\ref{T-matrix-ex}), $Q$ the
charge operator,
and \(\tau_{\pm}=(\tau_{1}\pm i\tau_{2})/\sqrt{2}\).  $M_1$ is the
constituent mass of the struck quark, and $M_2$ the mass of the
spectator antiquark,
which are solutions of the gap equation (\ref{gap}) due to the
dynamical chiral symmetry breaking.
In the case of pion, we set $M_1=M_2=M_{u(d)}$.  For $K^+$ meson,
$M_1=M_u$ and $M_2=M_S$.

We shall carry out the integration of (\ref{Tpion}) in the Bjorken
limit\cite{Lands,Shige};
%
%
\begin{eqnarray}
Q^{2}=-q^{2}\to\infty\; ,\; \; \nu=\frac{p\cdot q}
{m_{ps}}\to\infty\; ,\; \;
x=\frac{Q^{2}}{2m_{ps}\nu}\,:\, \mbox{fixed.} \nonumber
\end{eqnarray}
\vglue 0.3cm
\noindent
Here, $x$ is the so-called Bjorken $x$, and \(m_{ps}\) the pseudoscalar
meson mass.  We introduce the Sudakov variables;
%
%
\begin{eqnarray}
k_{\mu}=zp_{\mu}+yq_{\mu}+\kappa_{\mu} \; ,
\end{eqnarray}
\vglue 0.3cm
\noindent
where \(\kappa_{\mu}\) satisfies \(k\cdot p=k\cdot q=0\).
Thus, \(\kappa_\mu\) is spacelike \((\kappa^{2}< 0)\) and is
effectively two dimensional.
Calculating the traces and neglecting irrelevant terms in the Bjorken
limit\cite{Lands}, we obtain,
%
%
\vglue 0.02cm
\begin{eqnarray}
T_{\mu\nu}^{(-)}&=&\frac{8}{9}\frac{i}{(2\pi)^4}N_{c}g_{p qq}^2
\int\!dzd{\bar{y}}d^{2}{\kappa}
[t_{1}(\mu^{2},s)+zt_{2}(\mu^{2},s)]\frac{1}{z-x}\nonumber\\
& & \hspace {2cm} \times [-g_{\mu\nu}+\frac{2z}{m_{ps}\nu}
p_{\mu}p_{\nu}
+\frac{1}{m_{ps}\nu}(p_{\mu}q_{\nu}+p_{\nu}q_{\mu})]\;,
\label{TbeforeZ}
\end{eqnarray}
\vglue 0.3cm
\noindent
where
%
%
\begin{eqnarray*}
t_{1}(\mu^{2},s)&=&-\frac{1}{(\mu^{2}-M_{1}^{2}+i\varepsilon)^{2}}
\frac{1}{s-M_{2}^{2}+i\varepsilon}(\mu^{2}-M_{1}^{2})\\
t_{2}(\mu^{2},s)&=&-\frac{1}{(\mu^{2}-M_{1}^{2}+i\varepsilon)^{2}}
\frac{1}{s-M_{2}^{2}+i\varepsilon}(s+M_{1}^{2}-2M_{1}M_{2}-p^{2}) \; ,
\end{eqnarray*}
\noindent
and
%
%
\begin{eqnarray*}
\mu^{2}&=&(k-q)^{2}=z\bar{y}+{\kappa}^{2}\\
s&=&(k-q-p)^{2}=(z-1)(\bar{y}-p^{2})+\kappa^{2}\; .
\end{eqnarray*}
\vglue 0.2cm
\noindent
Here, we change the variable \(\bar{y}=2m_{ps}\nu(y-1)+zp^{2}\)
\cite{Lands}, and \(\mu^{2}\) and \(s\) are the invariant masses
of the struck quark and spectator.
Performing the z-integral in (\ref{TbeforeZ}),
we find\cite{Lands,Rujula},
%
%
\vglue 0.15cm
\begin{eqnarray}
T_{\mu\nu}&=&\frac{8}{9}\frac{1}{(2\pi)^3}N_{c}g_{p qq}^2
\int\! d{\bar{y}}d^{2}{\kappa}[t_{1}(\mu^{2}\, ,\,s)+zt_{2}
(\mu^{2}\, ,\,s)]\nonumber\\
& & \hspace {2cm} \times [-g_{\mu\nu}+\frac{2z}{m_{ps}\nu}p_{\mu}p_{\nu}
+\frac{1}{m_{ps}\nu}(p_{\mu}q_{\nu}+p_{\nu}q_{\mu})].
\label{TafterZ}
\end{eqnarray}
\vglue 0.3cm
\noindent
It is easily seen from (\ref{TafterZ}) that the calculated
structure functions exhibit the Bjorken scaling\cite{BJM}.
Consider now the \(\bar{y}\)-integral in the complex \(\bar{y}\)-plane.
The \(s\)-propagator has a pole at
\(\bar{y}=(\kappa^{2}-M^{2}_{2})/(1-x)+p^{2}+i\varepsilon/(1-x)\),
and the \(\mu\)-propagator has a double pole at
\(\bar{y}=(M^{2}_{1}-\kappa^{2})/x-i\varepsilon/x\).
For \(x > 1\) or \(x < 0\), all these singularities occur on the same
side of the real \(\bar{y}\)-axis,
and (\ref{TafterZ}) gives a zero result\cite{Lands,Rujula}.
However, if \(0 \leq x \leq 1\),
the integration over \(\bar{y}\) no longer
vanishes.  Integrating (\ref{TafterZ}) by \(\bar{y}\)
and taking the imaginary part, we get the quark distribution by
use of the optical theorem (\ref{dispersion}).
We change the integral variable $\kappa^2$ to \(\mu^{2}\), and find the
following expression for the quark distribution function.
%
%
\vglue 0.15cm
\begin{eqnarray}
q(x)&{\propto}&-g_{p qq}^{2} \int_{-\infty }^0 \! d\mu^{2}
[\frac{1}{\mu^{2}-M_{1}^{2}}   -x\frac{2M_{1}M_{2}-(M_{1}^{2}+
   M_{2}^{2})+p^{2}}{(\mu^{2}-M^{2}_{1})^{2}}]\nonumber\\
& & \hspace {3cm} \times \theta ({p}^{2}x(1-x)-xM_{2}^{2}-(1-x)\mu^{2})
\label{pi(x)}
\end{eqnarray}
\vglue 0.3cm
\noindent
Here, \(\theta\) is the usual step function which arises from
the spacelike condition of \(\kappa\).
We identify (\ref{pi(x)}) as the valence quark distribution of the
pseudoscalar meson \(q_{val}(x)\).  In the case of on-shell
pseudoscalar meson, we
use $p^{2}=m_{ps}^2$.  We use the Euclidean variable
$\mu_{E}^2 = -\mu^2$ for the integration of (\ref{pi(x)}) with the
relative momentum cutoff (\ref{cutoff}).
Note that the resulting distribution shows a
correct behavior $q(x) \rightarrow 0$ as $x \rightarrow 1$, since the
lower limit of the integral
$\, \mu _{E\min }^{2}(=-\mu^{2}_{\max})
={x \over {1-x}}M_2^2-xp^2 \rightarrow \infty \,$ as
$x \rightarrow 1$ and thus the integral vanishes\cite{BJM}.
We also note that the contribution of the second term of (\ref{pi(x)})
to the distribution function is small.  This smallness is due to the
spontaneous breakdown of the chiral symmetry.  In fact, the second term
disappears in the chiral limit; $m_u=m_d=m_s=0$.  This form ensures the
behavior $\, xq_{val}(x) \propto x \, $ at small $x$.
If the chiral symmetry were
not spontaneously broken, the second term would be as large as the
first term and the pionic
quark distribution would behave $ \, xq(x)_{val} \propto x^2 \, $
around small $x$.

We also apply the same procedure to the vector meson, though the NJL
model does not work well in the vector channel\cite{Vogl}.
Since the vector
mesons, e.g. $\rho$ meson, may be weakly coupled quark-antiquark
states, the
confinement of quarks is essential for the description of them.
Hence, it is not adequate to use the NJL model for the vector mesons
due to the absence of confinement in this model.  However, such an
attempt is still worth mentioning for an intuitive understanding of the
relation between the structure function and the quark interaction.

The hadronic tensor of the spin-1 vector meson is defined as,

\begin{eqnarray}
W_{\mu\nu}^{\alpha\beta}&=&[
-(g_{\mu\nu}-\frac{q_{\mu}q_{\nu}}{q^2})F_{1}(x)\nonumber\\
& &\hspace {1cm} + \, \frac{1}{m_{V}\nu}(p_{\mu}-\frac{p\cdot q}
{q^2}q_{\mu})  (p_{\nu}-\frac{p\cdot q}{q^2}q_{\nu})F_{2}(x)]\;
       (-g^{\alpha\beta}+\frac{p^{\alpha}
p^{\beta}}{p^2})
\end{eqnarray}
\vglue 0.1cm
\noindent
where $\alpha, \beta$ is the polarization of the vector meson, and
$m_{V}$ the vector meson mass.  This
form guarantees the current conservation, i.e.
$p_{\alpha} W_{\mu \nu }^{\alpha \beta }=
p_{\beta} W_{\mu \nu }^{\alpha \beta }=0$.  Contracting with the
meson polarization vector $\epsilon_{\alpha}$, $\epsilon_{\beta}$ and
summing over the helicity $\lambda$, we find the expression of the
unpolarized structure function;
\begin{eqnarray}
W_{\mu\nu} &=&\sum\limits_\lambda  {\varepsilon _\alpha (\lambda ,p)
^{*}\,\varepsilon _\beta (\lambda ,p)}\;
W_{\mu \nu }^{\alpha \beta }\nonumber\\
&=&(-g^{\alpha\beta}+\frac{p^{\alpha}p^{\beta}}{p^2})\;
W_{\mu \nu }^{\alpha \beta }\nonumber\\
&=&3[-(g_{\mu\nu}-\frac{q_{\mu}q_{\nu}}{q^2})F_{1}(x)\nonumber\\
& &\hspace{2.5cm}+\frac{1}{m_{V}\nu}(p_{\mu}-\frac{p\cdot q}
{q^2}q_{\mu})(p_{\nu}-\frac{p\cdot q}{q^2}q_{\nu})F_{2}(x)]\nonumber\\
\end{eqnarray}
\noindent
Corresponding to (\ref{Tpion}), $T_{-}^{\alpha\beta}$ in the
vector channel is given by,
\vglue 0.1cm
\begin{eqnarray}
T_{-}^{\alpha \beta}=S_{F}(k-q,M)g_{V qq}\tau_{+}\gamma^{\alpha}
S_{F}(k-q-p,M)g_{V qq}\tau_{-} \gamma^{\beta}S_{F}(k-q,M) \; .
\label{Trho}
\end{eqnarray}
\vglue 0.5cm
\noindent
Here, we omit the mass difference of the struck and spectator quarks
for simplicity.
After some trivial algebras, we obtain the quark distribution in the
vector meson.
\vglue 0.2cm
\begin{eqnarray}
q(x)&{\propto}&-g_{V qq}^{2}
\int {d\mu ^2}[{{1-x} \over {p^2}}-{2 \over {\mu ^2-M^2}}
+{{2x(p^2+2M^2)} \over {(\mu ^2-M^2)^2}}]\nonumber\\
& & \hspace {3cm} \times \theta (p^{2}x(1-x)-xM^{2}-(1-x)\mu^{2})
\label{rho(x)}
\end{eqnarray}
\vglue 0.5cm
\noindent
Here, $p^2 = m_V^2$.
We remark that the expression for the quark distribution in the vector
meson is quite different from that of the pseudoscalar meson.
In contrast to the pion case, the quark distribution function of the
rho meson behaves $ \, xq(x)_{val} \propto x^2 \, $ at the small $x$,
due to the large contribution of the third term in (\ref{rho(x)}).

Unfortunately, it was shown that the mass of $\rho$ meson is
more than two times
the constituent quark mass, and it should decay into two quark
states, within the standard parameters of the NJL model\cite{Vogl}.
Thus, we use a larger value for the constituent quark mass ($400MeV$)
and $G_V$ without changing other values
to obtain a 'bound state' for $\rho$ meson.

\vglue 3cm
\noindent
{\fontc {\bf 4 Numerical Results}}

In this section, we will show numerical results for quark
distribution functions with the use of the parameters in Table 1.
We first show in Fig.3 the quark distribution in the pion
(\ref{pi(x)}) at the
low energy model scale.  The peak of the resulting distribution
appears at $x \sim 0.6$, which indicates
asymmetric momentum distributions in the pion;
the struck quark carries a larger part of the pion
momentum.  This is due to large binding energy of the valence quark
in the pion.  Such a behavior is also suggested by the QCD
sum rule calculation\cite{QSR}.  This result is a consequence of
the highly non-perturbative structure of the pion

The valence quark distribution of kaon is also interesting.  We
can extract the effects of the explicit
$SU(3)$ flavor symmetry breaking
from the quark distribution of kaon.  The u- and s-quark distributions
of the kaon at the low momentum scale are presented in Fig.4.
The heavy strange quark carries a larger fraction of
kaon momentum than the light up(down) quark, as expected.

We note that this low energy scale structure function has no
physical meaning at this scale, since "real" structure function at the
low energy scale receives non-negligible contributions from all twist
operators.  The calculated results in Fig.3 and Fig.4
play roles of only the
boundary condition of the structure function for the \(Q^{2}\)
evolution.

We take the low energy hadronic scale at \(Q_{0}^{2}=(0.5 GeV)^{2}\),
which is used in ref. \cite{Gluck}.  At this scale, the running
coupling constant is still small;
\(\alpha_{s}(Q_{0}^{2})/\pi\sim0.3\).
Indeed, the inclusion of the second order QCD corrections gives a
small change
for the \(Q^{2}\) evolution from our result within 10\%\cite{Gluck}.

We may understand intuitively the physical meaning of this scale
\(Q_{0}^{2}\) as compared to the valence quark core radius of pion
\(\langle r^{2}\rangle _{core}\), as noted by Brown
{\em et al.}\cite{Brown}.
In their analysis of the pion electromagnetic form factor, the pion
consists of the valence quark core part, where a valence quark and
an antiquark move in the small region, and the sea quark
cloud part which is understood as the old vector meson dominance.
Their value is consistent to our low energy scale;
%
%
\begin{eqnarray*}
\langle r^{2}\rangle _{core}\sim (0.35 fm)^{2}\sim 1/(0.5GeV)^{2}\;\; .
\end{eqnarray*}
\vglue 0.3cm
We use the first order Altarelli-Parisi equation\cite{AP} for the
$Q^{2}$ evolution of valence distributions with
$\Lambda_{QCD}=250MeV$ to compare our results with experiment.
%
%
\vglue 0.1cm
\begin{eqnarray*}
{{dV(x,Q^2)} \over {d(\log Q^2)}}={{\alpha _s(Q^2)}
\over {2\pi }}\int_x^1 {dx}P_{qq}(x/y)V(y,Q^2)
\end{eqnarray*}
\vglue 0.3cm
\noindent
Here, $V(x,Q^2)$ is the valence quark distribution, and $P_{qq}$ the
splitting function\cite{AP}.

We show in Fig.5 the pionic quark distribution at $Q^{2}=20GeV^{2}$
by the solid curve with experimental data (the dashed curve)
extracted from the Drell-Yan process\cite{SMRS}.
We find a reasonable agreement with experiment.  The second moment
of the valence quark, which is identified with a momentum fraction
carried by the valence quark, turns out to be,

$ \hspace {1cm} <xu>_{\pi} = 0.22\hspace{1cm}$ at
$ \hspace {1cm} Q^{2} = 20GeV^{2}$  ,

\noindent
where $<xq> = \int_0^1 {dx}\,xq(x)$.  This value is remarkably
consistent with experimental data $0.21$\cite{SMRS}.
However, the calculated distribution function is almost zero at
$x \sim 1$, and different from the
experimental fit\cite{SMRS} or the counting rule
prediction\cite{Roberts}.  This shortcoming comes from
the cutoff procedure of the model.
Around $x\sim1$, the struck quark has a very large momentum $>
1 GeV$, and the quarks with so large momenta are excluded
in the NJL model by the cutoff.  At the moment,
such a high momentum quark can not exist in the hadron wave function
within the low energy quark model, and we ought to develop a model to
include the high momentum correlations consistently with the low energy
theory.

If we vary the low energy scale \(Q_{0}^{2}\) within 20\%,
the change of the quark distribution is rather small (\(\sim\)10\%).
As an example, we plot in Fig.6 the result in which we use
\(Q_{0}^{2}=0.7GeV^2\sim\Lambda_{NJLcut}^2\).  The peak position of the
calculated distribution is $x \sim 0.5$, which disagrees with
experiment.

We compare the pion structure function \(F_{\pi}(x)\) with experiment
in Fig.7.  Here, we do not take into consideration the sea quark
distributions, since we do not evaluate the sea quark diagrams at the
model scale.  It shows a good agreements for \(x > 0.3\).
In the low \(x\) region, the structure function is dominated by the
sea quark distributions.
As for the sea quarks, we can calculate their distribution functions
at the model scale as higher order loop corrections in the NJL
model\cite{Sea}.  We will discuss this point later.

The quark distributions in kaon are also evoluted to the experimental
scale in Fig.8.  The resulting
distributions show a similar flavor dependence with the previous
results obtained by the Regge theory
\cite{Regge1,Regge2}.
The corresponding second moments of the valence
quarks are given by,

$\hspace {1cm} <xu>_{K} = 0.20, \hspace {1cm} <xs>_{K} = 0.24
\hspace{1.5cm} $at
$ \hspace {1cm} Q^{2} = 20GeV^{2}$  .

\noindent
This result indicates that the heavy strange quark has a larger
momentum in kaon than the light u(d) quark.
These values are to be compared with the pion case; $<xu>_{\pi}= 0.22$.
The total momentum carried by the valence quarks in the kaon is
$<xu>_{K} + <xs>_{K} = 0.44$, and is almost the same as
that in the pion $ \, 2<xu>_{\pi} = 0.43$.  Similar results are
obtained by the recent QCD sum rule calculation\cite{Nishino}.

We also show in Fig.9 the ratio of kaon to pion valence u-quark
distributions $u_{K}/ u_{\pi} $ at $ Q^{2}=20GeV^{2}$.
The experimental values are taken from the Drell-Yan
experiment\cite{ratio}, and the analysis of the
large transverse momentum $\pi _0$ production processes\cite{frag}.
The result is consistent with available experiments.
Note that this ratio is sensitive to the mass difference of the
constituent quark masses in our model.
In fact, if we change the constituent strange quark mass,
the resulting ratio becomes quite different from experiment.
We plot two
cases ($M_s = M_{u}$ and $M_s = 2M_u$) in Fig.9 for comparison.

Finally, we show in Fig.10 the quark distribution in the rho meson at
the model scale.  This behavior is extremely different from the
result of the pion.  The resulting distribution shows a sharp peak at
$x\sim 0.5$ due to the weak correlation in the vector channel,
and resembles with the result of the non-relativistic static limit
 $q(x) \sim \delta(x-1/2) $.  Similar behavior is obtained in ref.
\cite{rhob1}.
Although the NJL model in the vector channel has some
difficulties\cite{Vogl}, the essential feature of this distribution
is expected to be model independent.

\vglue 3cm
\noindent
{\fontc {\bf 5 Semi-inclusive lepton nucleon scattering}}

Since a direct experiment of the lepton-pion deep inelastic scattering
is beyond the present experimental abilities, the information of
the pion structure function can be extracted from only the
$\mu$-on pair production process, as first suggested by
Drell and Yan\cite{DrellYan}.
Several authors considered an alternative approach to
determine the pionic quark distributions\cite{Semi1,Semi2,Semi3}.
In the semi-inclusive lepton nucleon scattering illustrated in Fig.11,
virtual meson clouds mainly contribute to this
process in the case of a slow nucleon in the final state.
This contribution is expected to be dominated by the one-pion exchange
from the study of the inclusive process $p+p\to n+X$, in which
the kinematical condition is similar with the semi-inclusive process
mentioned above; the proton beam carries the incident energy
$\sim 10 GeV$ and the momentum of the final state neutron is
$\sim 500MeV$\cite{ppscattering}.

It is also important to study such virtual pion processes in view of
the recent experimental observation,
i.e. 'the violation of the Gottfried sum rule'\cite{NMC}.
The pion clouds make substantial contributions to
the violation of the flavor symmetry in the nucleon sea
\cite{Henley,Kumano}.
In such a virtual pion lepton scattering, the current sees the
off-shell structure of the pion.  Here, we reexamine these processes by
taking into account the off mass shell dependence of the pion
structure function, which is neglected in the previous
works.

We consider the following process; (Similar expression can be
developed for the neutrino beam.)

$e+p\to e'+n+X$.

\noindent
The calculations are performed at the target rest frame with
the z-axis fixed by the current.
Following the work of Lusignoli {\it et al.},
we define kinematical variables shown in Fig.11\cite{Semi2}.
%
%
\vglue 0.05cm
\begin{eqnarray}
k& =&E_L (1,\sin \psi ,0,\cos \psi ) \hspace {1cm} \mbox{for the
incident electron} \nonumber\\
k'&=&E'_L (1,\sin (\psi +\theta _L ),0,
\cos (\psi +\theta _L )) \hspace {1cm}\mbox{for the outgoing electron}
\nonumber\\
P_{i}&=&M(1,0,0,0)
 \hspace {1cm} \mbox{for the target nucleon} \nonumber\\
P_{f}& =&(E,p\sin \alpha \cos \beta ,p\sin \alpha \sin \beta ,
p\cos \alpha )  \hspace {1cm} \mbox{for the outgoing nucleon}
\label{axis}
\end{eqnarray}
\vglue 0.1cm
\noindent
Here, $M$ is the nucleon mass, and $\theta _L$ the angle between
the electron momenta, and
$\tan \psi =E_{L}'\sin \theta _L / (E_L-E_{L}'\cos \theta _L)$,
which is chosen to provide $q=k'-k$ to have only the z-component.
$p$ is the three momentum of the final state
neutron.  The definition of the Bjorken variables are as follows;
\vglue 0.1cm
\begin{eqnarray}
\nu =q\cdot P_i/ M \; , \hspace{0.7cm}
Q^2=-q^2=4E_L E'_L \sin ^2{1 \over 2}\theta _{L} \;, \hspace{0.7cm}
x= Q^2 / 2M \nu  \;\;  .
\end{eqnarray}
\vglue 0.1cm
For the pion, we need to define its momentum and a quark momentum
fraction in the pion.
\begin{eqnarray}
P_{\pi} = P_i - P_f \, , \;\; t\equiv P_{\pi}^2 = 2M(M-E)\, , \;\;
x_{\pi} = Q^2 /2q \cdot P_{\pi}
\end{eqnarray}
\vglue 0.1cm
\noindent
It is easily shown that $x_{\pi}$ is related with the Bjorken $x$ as;
\vglue 0.05cm
\begin{eqnarray}
{x \over {x_\pi }}={{q\cdot P_\pi } \over {q\cdot P_i}}=1-{E \over M}+
{{\sqrt {\nu ^2+Q^2}} \over \nu }{p \over M}\cos \alpha \;\; .
\end{eqnarray}
\vglue 0.1cm
The requirement of the kinematics implies the following conditions
with $\nu \rightarrow \infty$,
\vglue 0.05cm
\begin{eqnarray}
x\le x_{\max} \sim 1-(E-p)/ M  \le 1 \; , \hspace {1cm}
\cos \alpha \ge {E \over p}-{M \over p}(1-x) \; .
\end{eqnarray}
\vglue 0.1cm
\noindent
The total (unobserved) missing mass $M_{X}$
of this process is given by,
\vglue 0.05cm
\begin{eqnarray}
M_{X}^2 = (q+P_{\pi})^2 = Q^2 (\frac{1}{x_\pi} - 1) + t \; .
\end{eqnarray}
\vglue 0.1cm
\noindent
The condition of the deep inelastic scattering also requires the
sufficiently large missing mass.  In the present study, we assume that
$M_{X}$ is larger than $2GeV$; $M_{X \min} = 2GeV$.
Thus, the variable $x_{\pi}$ is constrained by the above conditions as;
%
%
\vglue 0.1cm
\begin{eqnarray}
x \le x / x_{\max}\le x_\pi \le
\frac{Q^2} {Q^2 - t + M_{X \min}^2} \le 1  \;\; .
\label{piregion}
\end{eqnarray}
\vglue 0.1cm

The cross section of the electron-proton process is calculated
as\cite{Semi2},
%
%
%
\vglue 0.1cm
\begin{eqnarray}
\frac{d\sigma}{dE'_{L}d \cos \theta_{L}d^{3}p} &=&
     \frac{4\alpha^{2}}{\pi Q^{4}} \frac{g^{2}_{\pi NN}}{4\pi}
     \frac{E'^{2}_{L}}{ME} \frac{-t}{(t-m_{\pi}^{2}) ^ 2}\nonumber\\
        & & \times \{
 2 \sin^{2}\frac{1}{2}\theta_{L}\,F_{1}^{\pi}(x_{\pi},Q^2 , t)
\nonumber\\
 & &+[\frac{(k\cdot P_{\pi})(k'\cdot P_{\pi})}{E_{L}E'_{L}}
 -t \sin^{2}\frac{1}{2}\theta_{L}]
        \frac{F_{2}^{\pi}(x_{\pi},Q^2 , t)}{q \cdot P_{\pi}}
        \} \; ,
\label{cross}
\end{eqnarray}
\vglue 0.3cm
\noindent
\noindent
where
%
%
\begin{eqnarray}
        F_{1}^{\pi}(x_{\pi},Q^2 , t) =
        \frac{1}{2x_{\pi}}F_{2}^{\pi}(x_{\pi},Q^2 , t)
\label{offpi}
\end{eqnarray}
\vglue 0.2cm
\noindent
is the pion structure functions. $m_{\pi}$ is the pion mass, and
$g_{\pi NN}=13.5$ the pion-nucleon coupling constant.

For the pion exchange mechanism, we examine
the elementary pion with
the dipole ${\pi}NN$ form factor and the Reggeized pion.
In the former case, the pion-nucleon coupling becomes,
\begin{eqnarray}
g_{\pi NN} \Rightarrow
g_{\pi NN} \frac {(1 - m_{\pi}^2 / {\Lambda}^2 )^2}
{(1 - t/{\Lambda^{2}})^2}
\label{dipole}
\end{eqnarray}
\vglue 0.5cm
\noindent
where $\Lambda = 1GeV$ is the momentum cutoff of the
$\pi NN$ interaction.
On the other hand, in the latter case, we replace the pion propagator
square with the triple Regge formula\cite{Semi2}.
%
%
\vglue 0.1cm
\begin{eqnarray}
\frac{1}{(t-m_{\pi}^{2}) ^ 2} \Rightarrow \pi^{2}\alpha'^{2}
       \frac{1+\cos [\pi\alpha_{\pi}(t)]}
	   {2\sin^{2}[\pi\alpha_{\pi}(t)]}
       e^{b(t-m_{\pi}^{2})}
[\frac{1/x_{\pi} -1 -t/q^2 }{1/x - 1 - M^2 /q^2}]^{-2\alpha_{\pi}(t)}
\label{tregge}
\end{eqnarray}
\vglue 0.5cm
\noindent
Here, $\alpha _\pi (t)=\alpha '(t-m_{\pi}^2)$ with
$\alpha '= 1GeV^{-2}$.
Exponential damping factor $b$ is chosen phenomenologically to
reproduce the data of $p+p\to n+X$ scattering; $b= 0.56 GeV^{-2}$.

We first show in Fig.12 the off-shell $t$-dependence of the pion
structure function evoluted to $Q^2 = 20GeV^2$, using (\ref{pi(x)}),
where $p^2 = t$ and $g_{\pi qq} (t)$.  Here, $g_{\pi qq}(t)$
is almost unchanged from the on-shell case.
This effect is neglected in the previous works\cite{Semi2}.
As the pion momentum $t$ increases,
the peak position of the quark distribution moves toward the small $x$
region, and the distribution function shows a substantial reduction
for $x>0.4$.
Around $x \sim 0.5$, the absolute value of the
distribution function at $t=-0.5GeV^2$ is almost half of that for the
on-shell case.
This decrease is crucial to understand the following results.

Using (\ref{cross}) with (\ref{dipole}) or (\ref{tregge}),
we obtain the cross section with the incident
electron energy $E_{L}=30GeV$.
Here, we use the distribution functions at $Q^2 = 20 GeV^2$ shown
in Fig.12 for simplicity, since the change of distributions due to the
$Q^2$ evolution is rather small ($\leq 20 \%$) in the momentum region
above $Q^2=5GeV^2$, which is about the same size as the experimental
uncertainty of the pion structure function\cite{Kumano}.

For comparison, we also present the results with other
plausible forms of
the pion structure function; the experimental fit of Sutton
{\it et al.}\cite{SMRS}, and the parameterization of Gl{\"u}ck
{\it et al.}\cite{Gluck}.
These are chosen to reproduce the on-shell data
of the pionic distribution, that is, they do not depend on
the pion momentum.  In each figure, we show our calculation by the
solid curve.  The results with the distribution functions of
Sutton {\it et al.} and Gl{\"u}ck {\it et al.}
are depicted by the dashed and dotted curves, respectively.

Concerning the sea quark distribution, we do not calculate it at the
model scale, and thus we can not estimate its contribution.
Experimental determination also contains a
large uncertainty\cite{SMRS}.
Hence, we simply use the parameterization of Gl{\"u}ck
{\it et al.}\cite{Gluck} for the sea quark distributions in all
the cases.  The sea quark contribution to the cross
section is negligible for $x>0.3$.

We show in Fig.13 $(d\sigma/dp)$ with $p$ being the three momentum
of the neutron in the final state.
We utilize the triple Regge formula (\ref{tregge}) in
Fig.13 (a), and the dipole vertex (\ref{dipole}) in Fig.13 (b).
The results depend on the type of the
$\pi NN$ vertex, especially $p > 0.3GeV$, whereas they are almost
independent of the choice of the structure function, i.e. the off-shell
effect is not evident.  Thus, we can get the most suitable form
of the $\pi NN$ exchange process from these figures by comparing with
the forthcoming experimental data.

We next discuss the $x$ dependence of the cross
section $(d\sigma / dx)$ shown in Fig.14 (a) and (b).
For $x > 0.3$, the resulting cross section shows a clear difference
between ours and others.
The result with the off-shell structure function is much
smaller than that with the experimental fit for $x>0.3$.
Note that the sea quark contribution is very small for $x>0.3$.
Hence, this difference is caused purely by the change of the valence
quark distribution function.  It is understood as a convolution effect.
Roughly speaking, the cross section at $x$
is expressed as an integral of the pion structure function $F_{\pi}
(x_{\pi})$.  Because of (\ref{piregion}),
this integration over $x_{\pi}$ is essentially carried out from
$x$ to 1, namely, the partons with the
momentum fraction $x_{\pi} \geq x$
in the pion do contribute to the cross
section at $x$.  Thus, the cross section is more sensitive
to the change of the large $x_{\pi}$ behavior of the distribution
function.
We recall that the reduction of the calculated off-shell
distribution function from the on-shell one becomes
large at the large $x$ region shown in Fig.12.
Therefore, as $x$ increases, the difference of the two cross
sections becomes considerably large.

Unfortunately, all of this decrease is not originated from the
off-shell effect.  To see this point, we also present the result by the
dash-dotted curve in Fig.14, using the calculated pion structure
function with no off-shell dependence, i.e. we neglect the
$t$-dependence of the distribution function and fix
$t=m_{\pi}^2=(140MeV)^2$.  It gives slightly smaller cross section than
those of Sutton {\it et al.} and Gl{\"u}ck {\it et al.} for $x>0.3$.
This reduction comes from the discrepancy between our calculation and
the
empirical distribution function for the on-shell pion, already shown
in Fig.5.  Because our quark distribution is different from the
empirical one at $x \sim 1$, this disagreement reduces the cross
section.  However, our two results, $t$-dependent case and
$t=m_{\pi}^2$
fixed case, also show a clear difference in Fig.14.  This reduction of
the cross section is due to the real off-shell effect, as discussed
above.
\vglue 3.0cm

\noindent
{\fontc {\bf 6 Summary and Discussions}}

In this paper, we have studied the deep inelastic structure function
of mesons
using the Nambu and Jona-Lasinio model as a low energy effective model
of QCD.  The calculations of the leading twist operator are performed
at the low energy model scale $Q^2_0$, and the resulting distributions
are evoluted to the experimental large momentum scale in terms of the
Altarelli-Parisi equation.
Although next to leading order contributions of
the QCD evolution are small\cite{Gluck},
it is very difficult to estimate
uncertainties for the use of the QCD perturbation below $1GeV$.

Our results are in reasonable agreements with experimental data,
except for the large $x$ region.
In this region, the struck quark carries a large
momentum $> 1GeV$, and the NJL model is not designed for the momentum
$p^2 > \Lambda^2 \sim 1GeV^2$.  Generally, the phenomenological
quark wave functions, e.g. Isgur-Karl model(Harmonic oscillator type)
or the MIT bag model, do not include such high momentum components.
We should
improve the behavior of the structure function at the large $x$, by
taking into account the quark correlation in the high momentum region.
Higher twist contributions are also expected to change the shape of the
distribution function at the large $x$.  Indeed,
experimental analysis of the scaling violation
indicates that higher twists give non-negligible contributions to the
nucleon structure functions in the large $x$ region.
If the pion is a collective ${q\bar{q}}$ state, twist-4
contributions may be larger than those of the nucleon.

The extraction of the flavor symmetry breaking effects in DIS is also
important.  By studying the differences between quark distributions of
various flavors, we obtain the information about the flavor
dependence of their internal wave functions in hadrons,
e.g. the nucleon structure function gives the spin-flavor
structure of the valence quarks for the u- and
d-quarks\cite{Diquark}.  The kaon structure function
provides the valuable data on the strange sector.
As we have discussed for the kaon, the valence strange quark may carry
larger momentum fraction than the up or down quark in the kaon.
Comparing the u-quark distribution in the kaon $u_{K}(x)$ with that in
the pion $u_{\pi}(x)$, the NJL model calculation indicates the
dominance of  $u_{\pi}(x)$ at the large $x$, and is consistent with the
available data.  This is due to the strong quark correlation in
the pion, namely, the binding energy of the u-quark in the pion is
larger than that in the kaon.

The investigation of the semi-inclusive lepton nucleon scattering is
very interesting as an alternative method to extract the virtual meson
contributions to the nucleon structure.  From them, we can
obtain the
off-shell dependence of the quark distribution function in the
virtual pion.  It
also helps deeper understandings of the $SU(2)$ flavor symmetry
breaking in the nucleon sea.  The calculated distribution function
shows rather large momentum dependence.  The pion structure function
decreases as the spacelike momentum of the pion $t$
becomes large, particularly for $x>0.4$.
We have computed the cross section with the use of
the off-shell pion structure
function and the on-shell one, and found the cross section of the
off-shell $t$-dependent
case is much smaller than the $t=m_{\pi}^2$ fixed case for $x>0.3$.
In order to compare our calculations with the forthcoming experiment,
we ought to study this process more
rigorously with the improvement of the large $x$ behavior of the
distribution function, e.g. estimation of contributions from other
mesons and nucleon resonances, or inclusion of the interference of the
${\pi}$ and ${\eta}$ meson\cite{Semi3}.
Such a study is under investigation\cite{preparation}.

Here, we comment on the sea quark distributions in the NJL model.
The lowest order diagram for the sea quark distributions is shown in
Fig.15 (a), corresponding to Fig.1 (a), which is very important for
the calculation of the quark self-energy and hence the spontaneous
chiral symmetry breaking.
As discussed by Bernard {\it et al.}\cite{BJM}, however,
this sea quark diagram does not contribute to the structure function
in the NJL model, since it has no imaginary part.
In order to get the 'meaningful' sea quark distribution in the model,
we must calculate next to leading order diagram shown in Fig.15 (b),
which are usually neglected within the Hartree approximation.
This inconsistency of the approximation is a defect of
the NJL model.  It can be shown that the contribution of the next to
leading order diagram to the quark self-energy is about 10\% of the
total constituent mass.  Therefore, the sea quark distribution is
expected to share the momentum fraction of the same magnitude in the
pion at the low energy model scale\cite{Oursea}.

Our approach in terms of the NJL model achieves
qualitative agreements with the existing experimental data.
The calculated
quark distributions reflect the low energy quark dynamics of the model,
i.e. the single-particle energy of the valence quark
and the quark correlation
in mesons.  Such agreements may indicate the possible connection
between the low energy quark model and the deep inelastic phenomena.

We thank Prof. Y. Mizuno for valuable discussions about the
semi-inclusive electron nucleon scattering.

\newpage
{ \bf Table 1}  Meson Properties

\noindent
Parameters

\noindent
$m_{u,d}=5.5MeV, \, m_{s} = 130MeV, \, \Lambda = 900MeV, \\
\, G_{S}\Lambda^2 = 2G_{V}\Lambda^2 = 19, \, a=0.15GeV^{2}$
\vglue 0.5cm

\begin{tabular}{lcc}
& & \\ \hline
  & Theoretical values & Empirical values \\ \hline
Pion mass & 140MeV &140MeV \\
$f_{\pi}$ & 93MeV & 93MeV \\
Kaon mass & 498MeV & 495MeV \\
$f_{K}$& 96MeV & 114MeV \\
$M_u$& 370MeV & \\
$M_s$& 548MeV & \\
$<\bar \psi \psi>$& $(-250MeV)^3$ & $\sim (-250MeV)^3$\\
$g_{\pi qq}$& 3.92 & \\  \hline
\end{tabular}

%
%
%
\newpage
\noindent
{\bf Figure Captions\/}

\ni
Fig. 1 (a)	: The Dyson equation for the quark propagator.  The thick
solid line represents the dressed constituent quark,
and the thin solid line the current quark.
\vspace {0.5cm}

\ni
Fig. 1 (b)	: The Bethe-Salpeter equation for quark-antiquark
scattering.
${\cal T}$ is the quark-antiquark T-matrix with the total momentum
$q$.  ${\cal K}$ denotes the interaction kernel discussed in the text.
\vspace {0.5cm}

\ni
Fig. 2 : The forward scattering amplitude ("handbag diagram") of the
pion.  The solid line represents the quark, and dashed line
the meson.  The virtual photon is depicted by the wavy line.
For details and notation, see text.
\vspace {0.5cm}

\ni
Fig. 3	: The quark distributions of the pion at the low energy model
scale \(Q^{2}=Q_{o}^{2}\) as a function of the Bjorken \(x\).
\vspace {0.5cm}

\ni
Fig. 4	: The quark distributions of the kaon at the low energy model
scale \(Q^{2}=Q_{o}^{2}\).  The u-quark distribution is depicted by
the solid curve, and the s-quark distribution by the dashed curve.
\vspace {0.5cm}

\ni
Fig. 5	: The valence quark distribution of the pion at
$Q^2=20GeV^2$ (solid curve) as a function of the Bjorken $x$,
in which we use the model scale $Q_{0}^{2}= 0.25GeV^2$.
The experimental fit\cite{SMRS} is depicted by the dashed curve.
\vspace {0.5cm}

\ni
Fig. 6	: The valence quark distribution of the pion at
$Q^2=20GeV^2$ with the model scale $Q_{0}^{2}= 0.7GeV^2$.
The notations are the same as those in Fig.5.
\vspace {0.5cm}

\ni
Fig. 7	: The pion structure function at $Q^2=20GeV^2$.
The experimental data are taken from the NA3 experiment\cite{DrellYan}.
The theoretical prediction of the NJL model is depicted by
the solid curve.
Here, we use $F_{\pi}(x)=Kxq(x)$ with the sea quark distributions
being neglected, where we take the $K$-factor, $K=1.5$\cite{SMRS}.
\vspace {0.5cm}

\ni
Fig. 8	: The valence quark distributions in the kaon at
$Q^2=20GeV^2$ with the model scale $Q_{0}^{2}= 0.25GeV^2$.
The u-quark distribution is depicted by the solid curve, and the
s-quark distribution by the dashed curve.
\vspace {0.5cm}

\ni
Fig. 9	: The ratio of kaon to pion valence u-quark distributions
$u_{K}(x)/u_{\pi}(x)$ at $Q^2=20GeV^2$.
The theoretical result is depicted by the solid curve.
The closed circles with error bars are taken from the Drell-Yan
experiment\cite{ratio}.  The open circle line with the dotted area,
which indicates the error, is obtained by taking a ratio
of the parameterization in ref. \cite{frag}, where the pion and kaon
structure functions are determined by the data on the
large transverse momentum $\pi _0$ productions.
For comparison, the results with $M_{s}/M_{u}=$2 and 1 are shown
by the dashed and dotted curves, respectively.
\vspace {0.5cm}

\ni
Fig. 10	: The valence quark distribution in the rho meson at the model
scale as a function of the Bjorken $x$.
\vspace {0.5cm}

\ni
Fig. 11	: The semi-inclusive lepton nucleon scattering.  The solid line
represents the electron and the doubled line the nucleon.  The virtual
pion and the virtual photon are depicted by the dashed and wavy lines.
\vspace {0.5cm}

\ni
Fig. 12	: The off-shell dependence of the quark distribution in the
pion.  The results at
$t=m_{\pi} ^2$ (on-shell), and $t=-0.1, -0.25, -0.5 GeV^2$ are shown.
\vspace {0.5cm}

\ni
Fig. 13	: $d{\sigma}/dp (e+p \rightarrow e' + n + X)$ as a function
of $p$ with $E_{L} = 30GeV$.
In Fig.13 (a), the triple Regge formula is used for
the ${\pi NN}$ vertex, and  the dipole cutoff in Fig.13 (b).
The theoretical result is depicted by the solid curve.
The results using the distribution of Sutton {\it et al.}\cite{SMRS}
and Gl{\"u}ck {\it et al.}\cite{Gluck} are also shown by the
dashed and dotted curves, respectively.
\vspace {0.5cm}

\ni
Fig. 14	: $d{\sigma}/dx (e+p \rightarrow e' + n + X)$ as a function
of the Bjorken $x$ with $E_{L} = 30GeV$.
The triple Regge formula is utilized in Fig.
14 (a), whereas the dipole formula in Fig.14 (b).   The notations are
the same as those in Fig.13.  The result in terms of the calculated
structure function with no off-shell dependence is also shown by the
dash-dotted curve.
\vspace {0.5cm}

\ni
Fig. 15(a) : The lowest order sea quark diagram in the NJL model.  The
solid line denotes the quark, and the wavy line the virtual photon.
\vspace {0.5cm}

\ni
Fig. 15(b) : The next to leading order sea quark diagram in the NJL
model.  The dashed line indicates the pion.
\vspace {0.5cm}
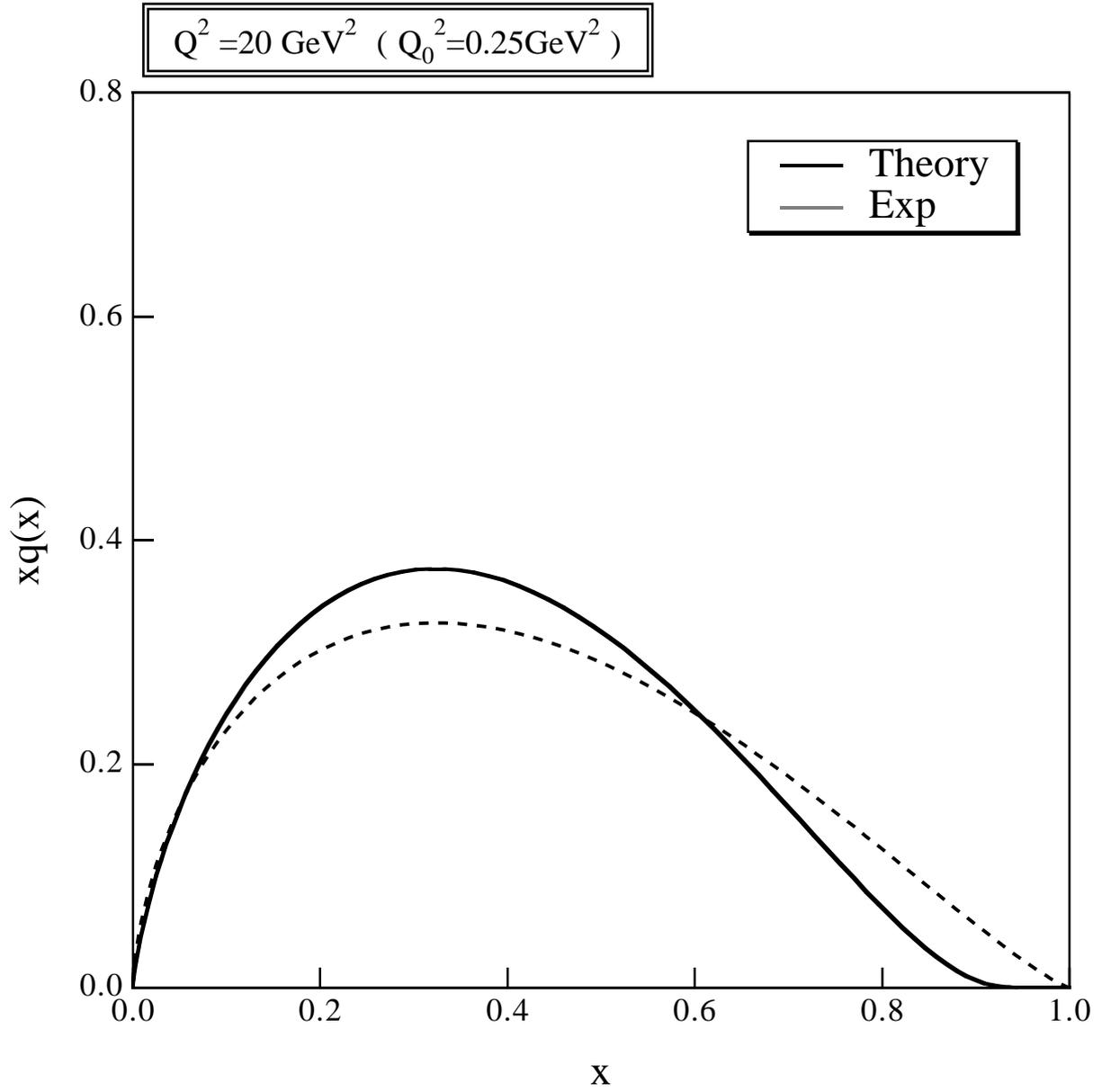
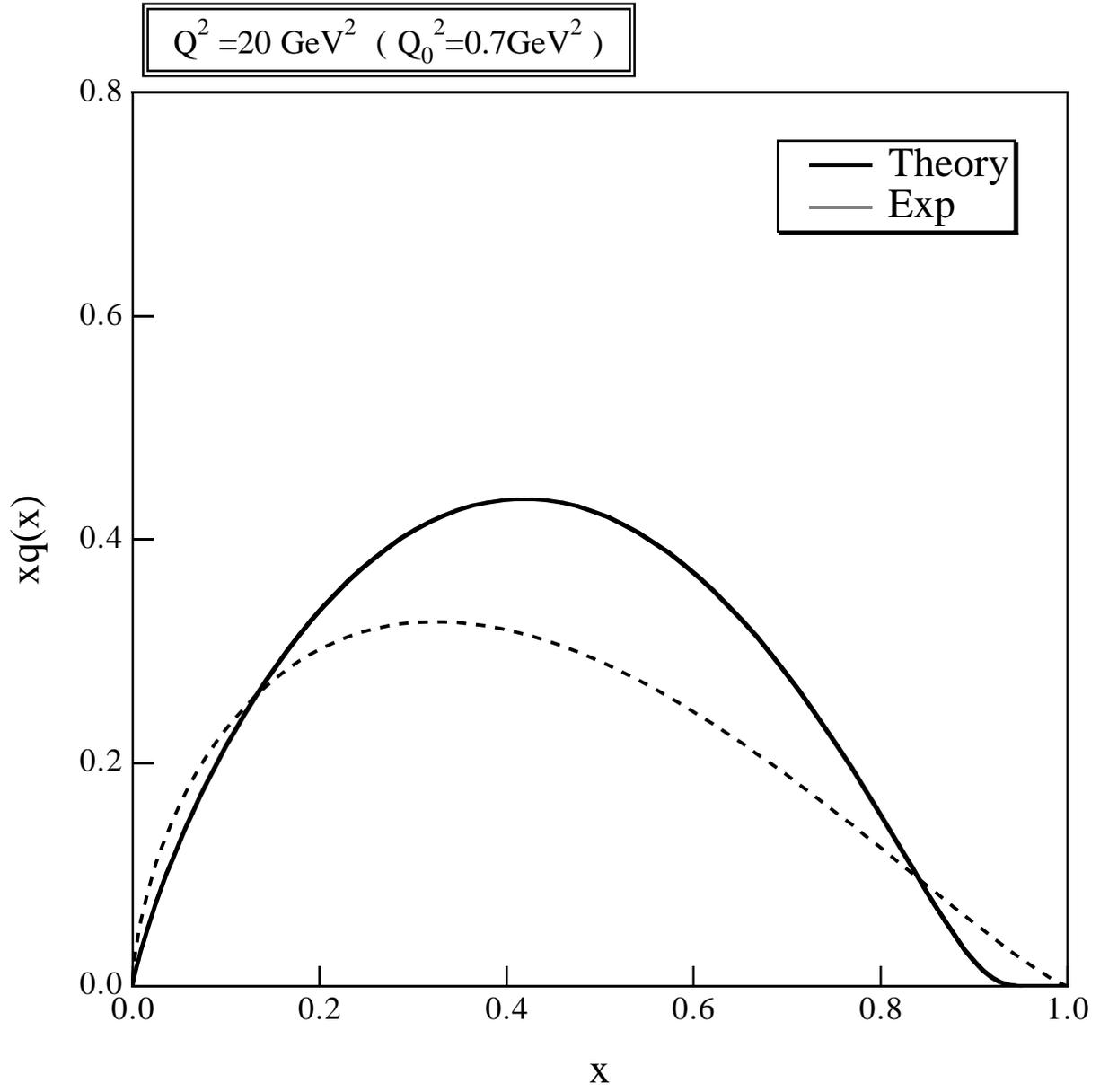
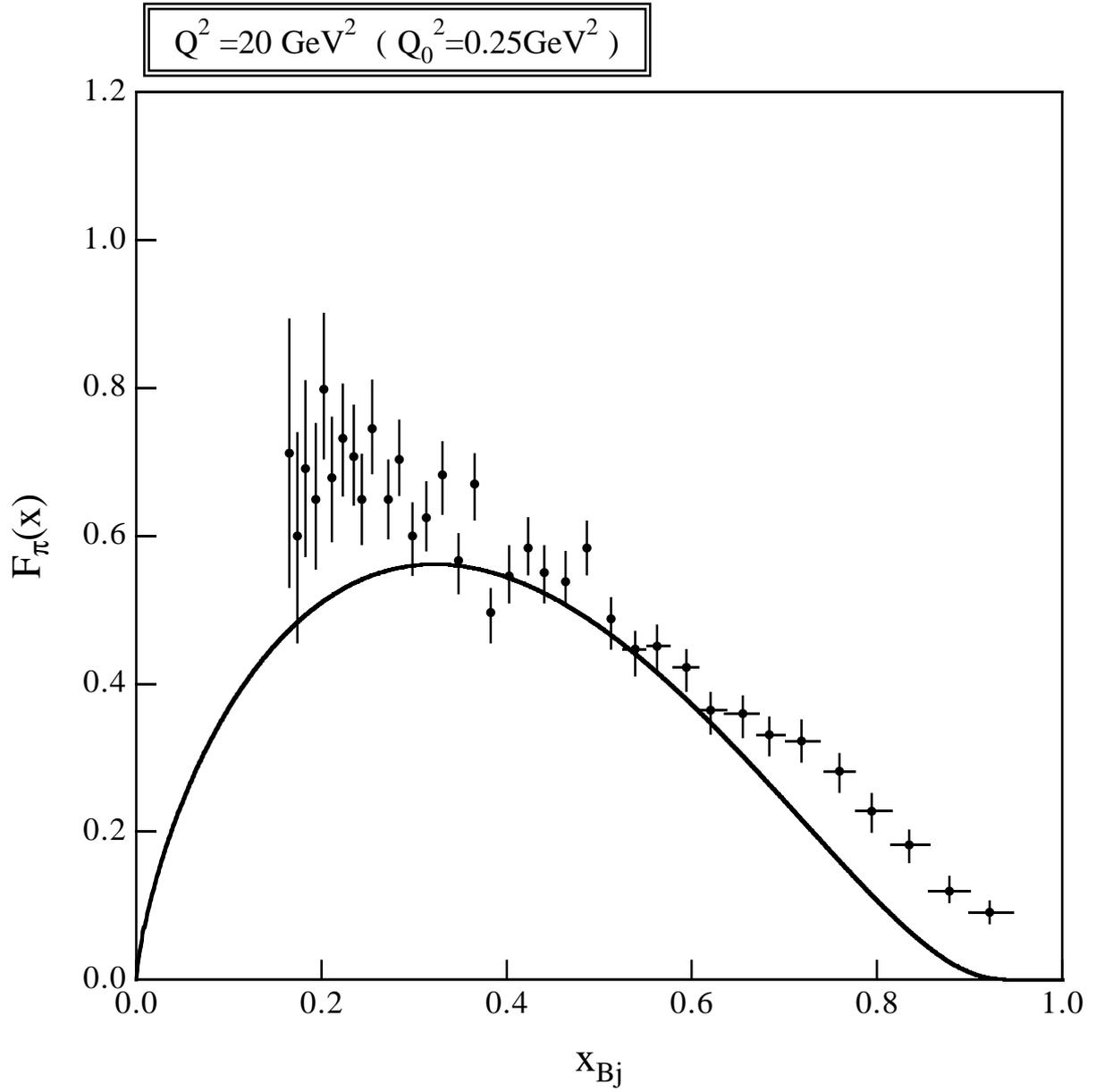
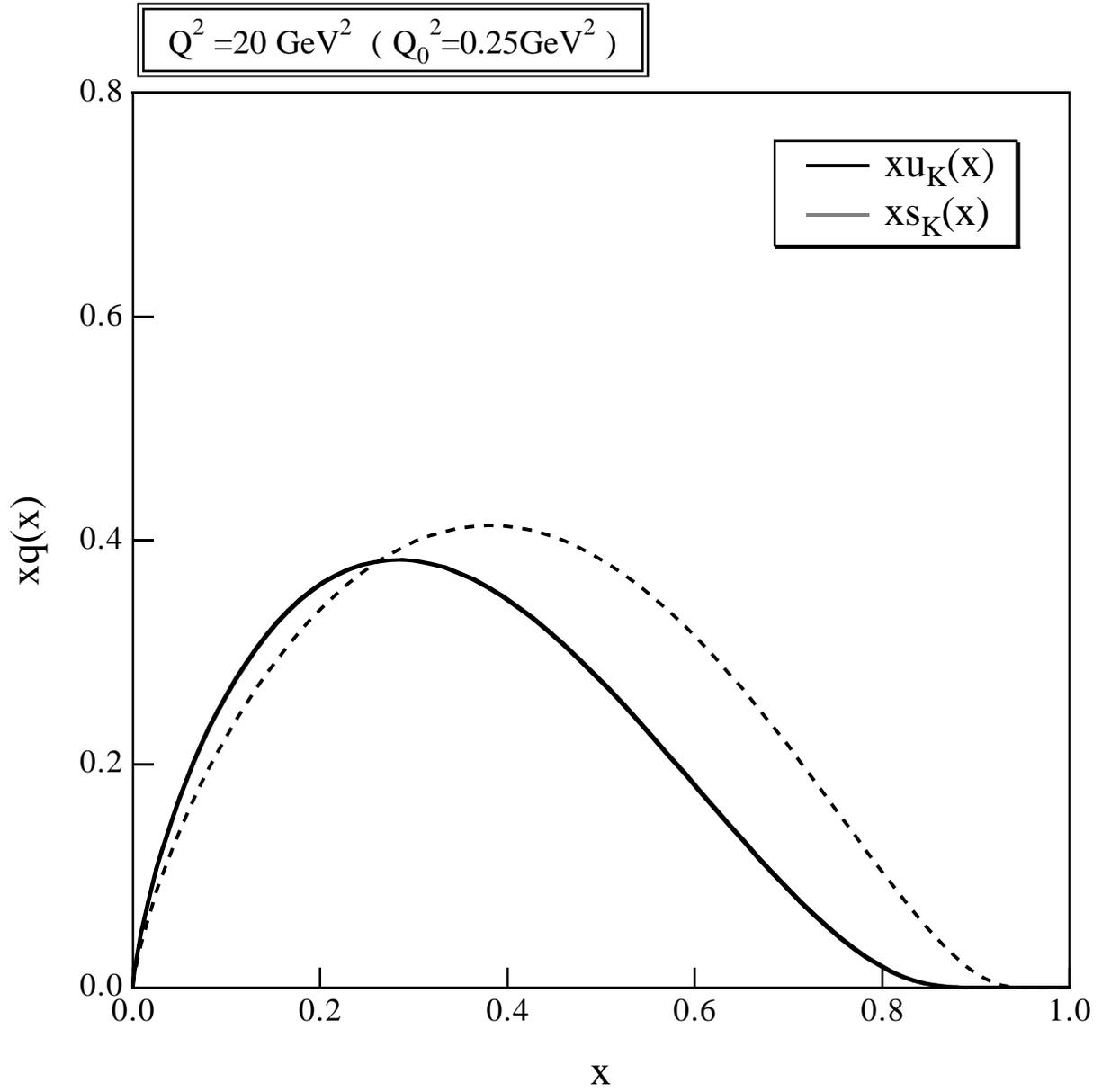
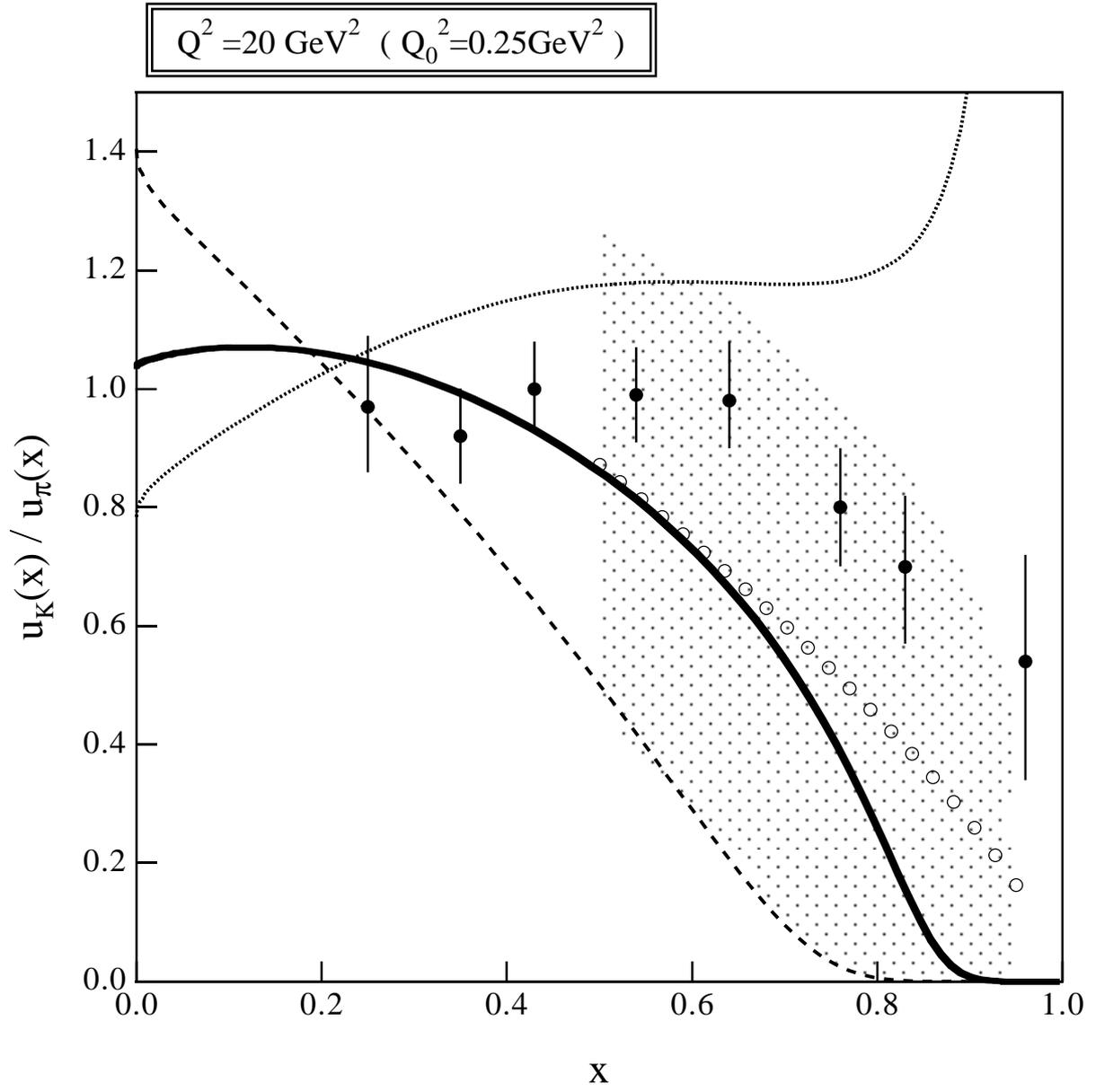
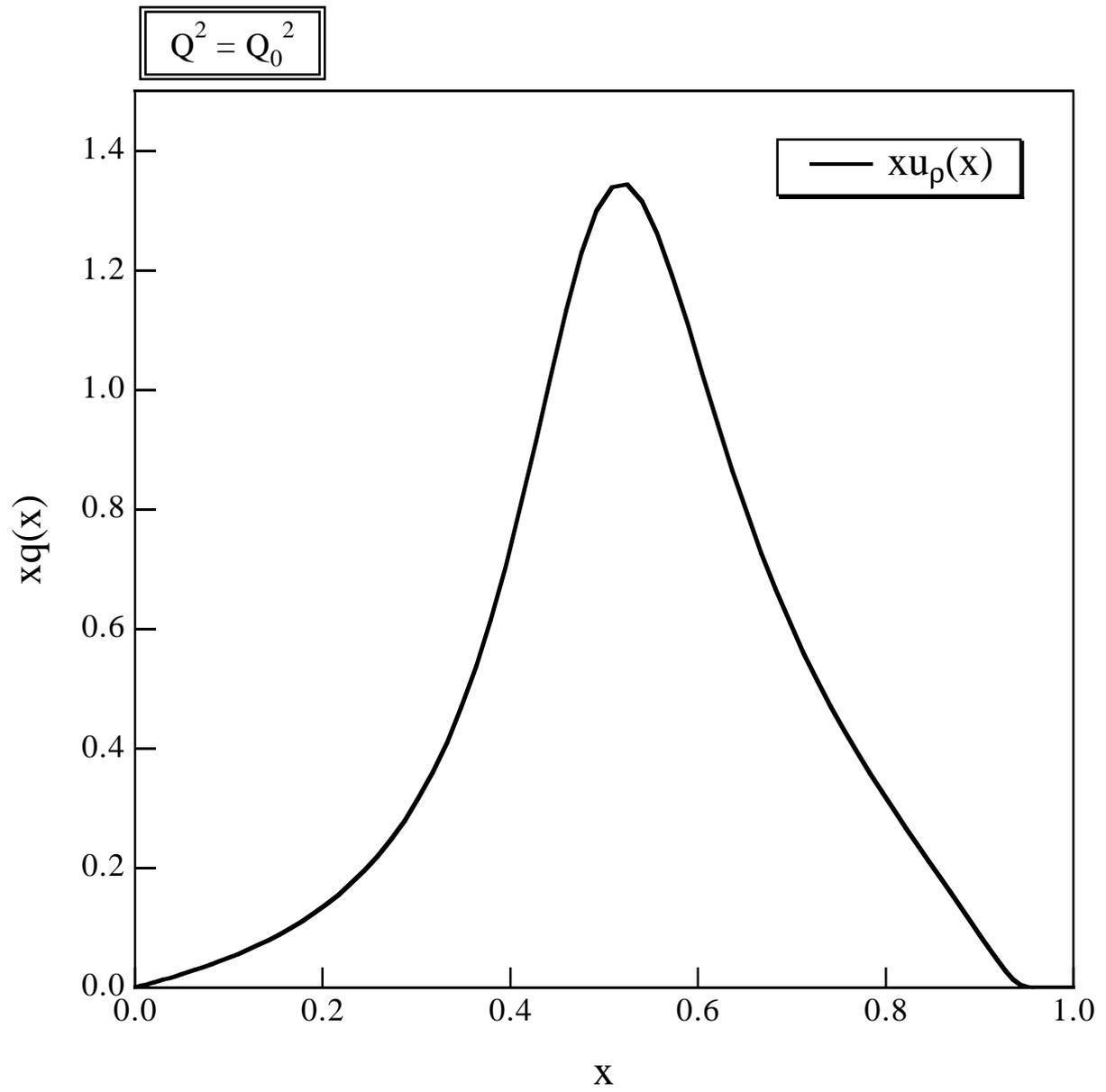
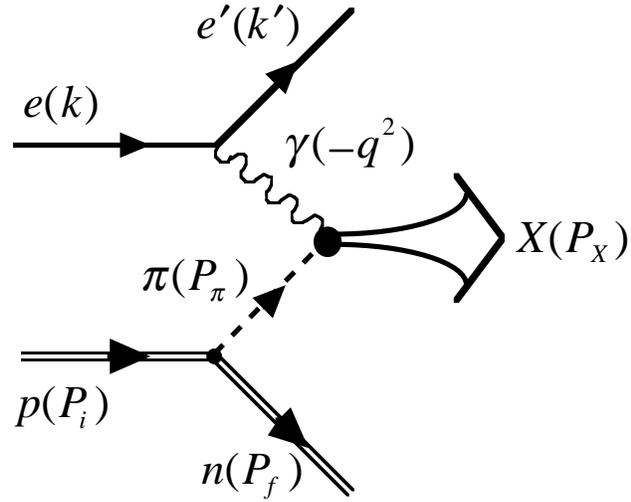
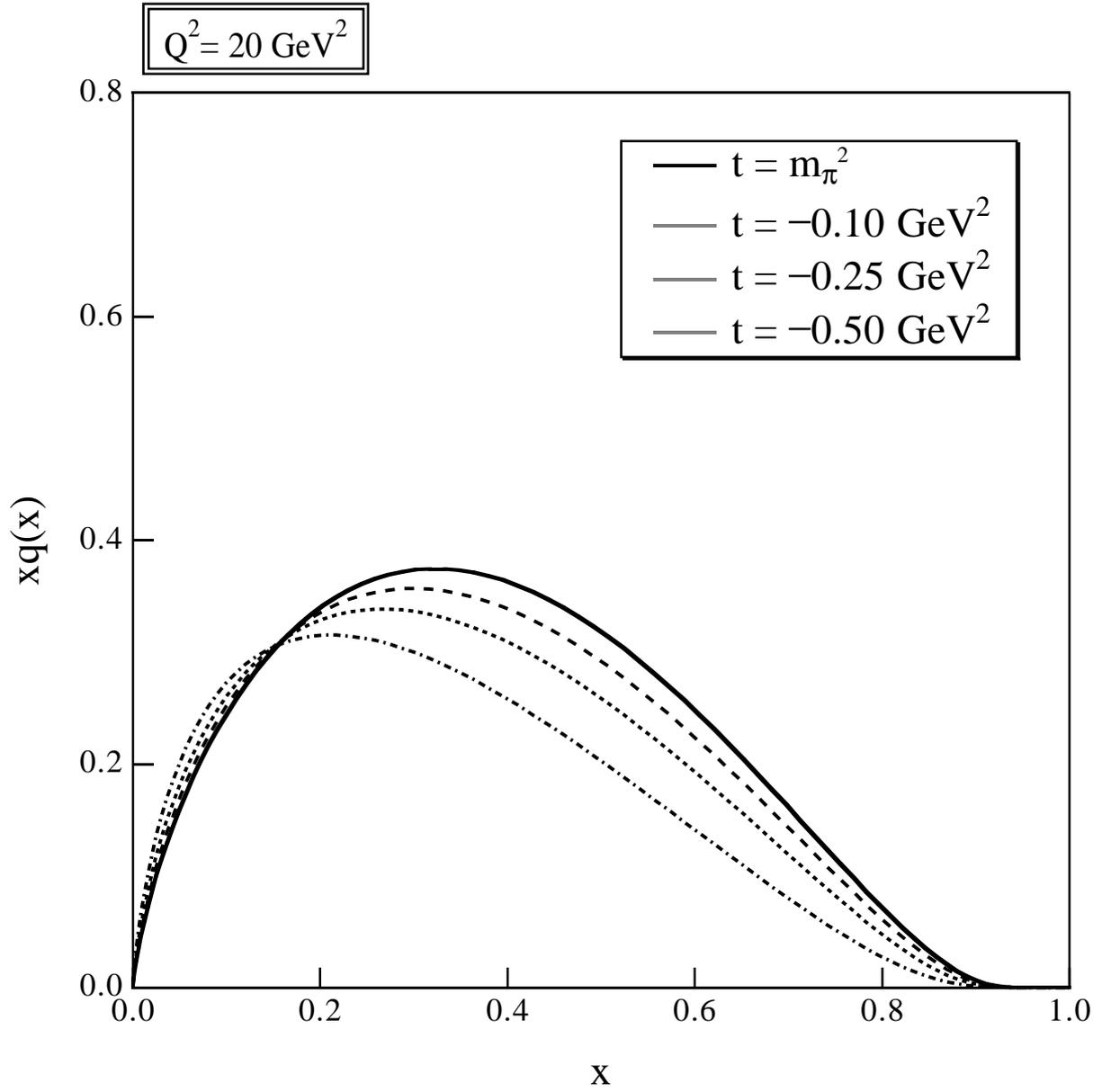
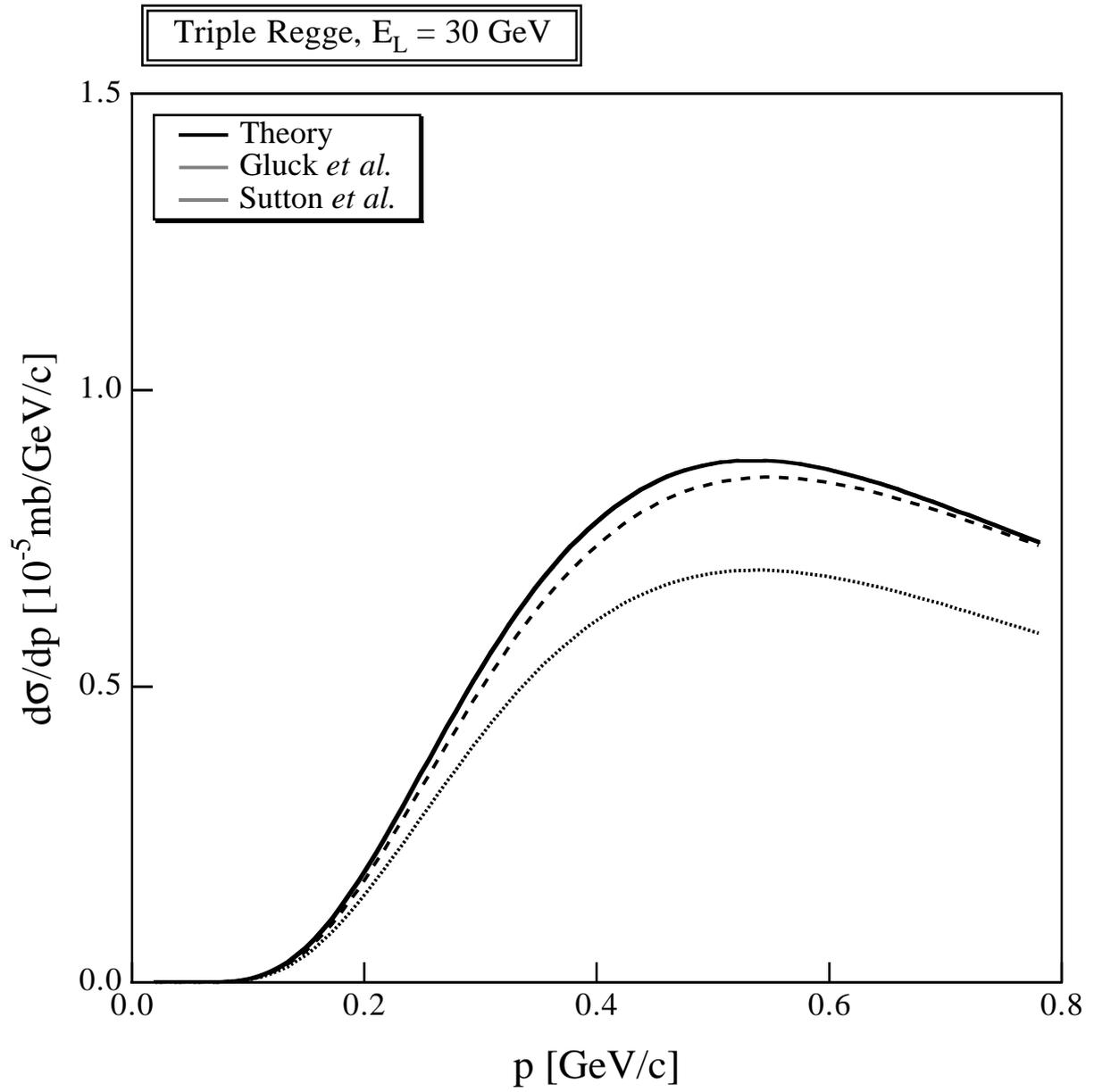
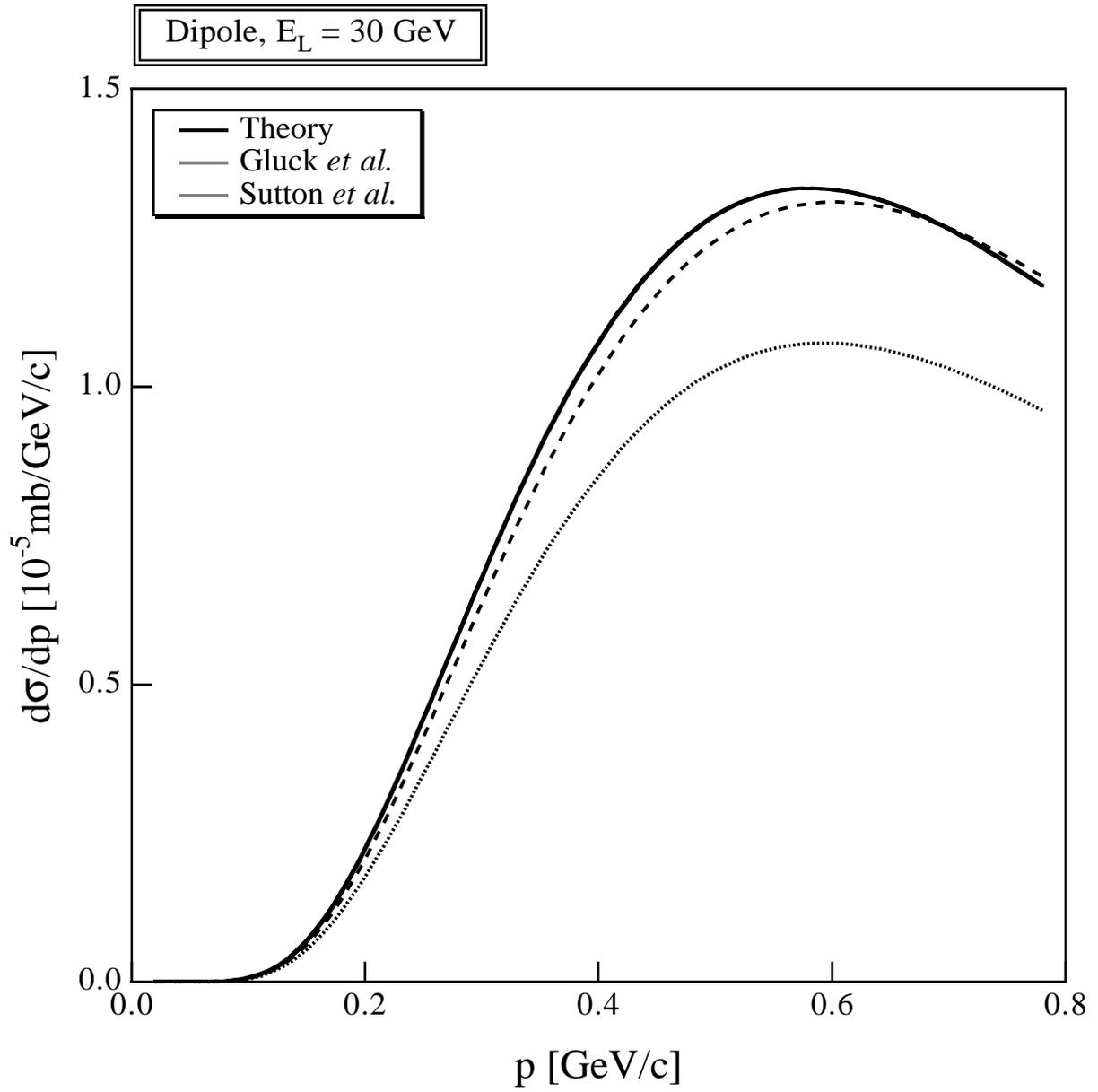
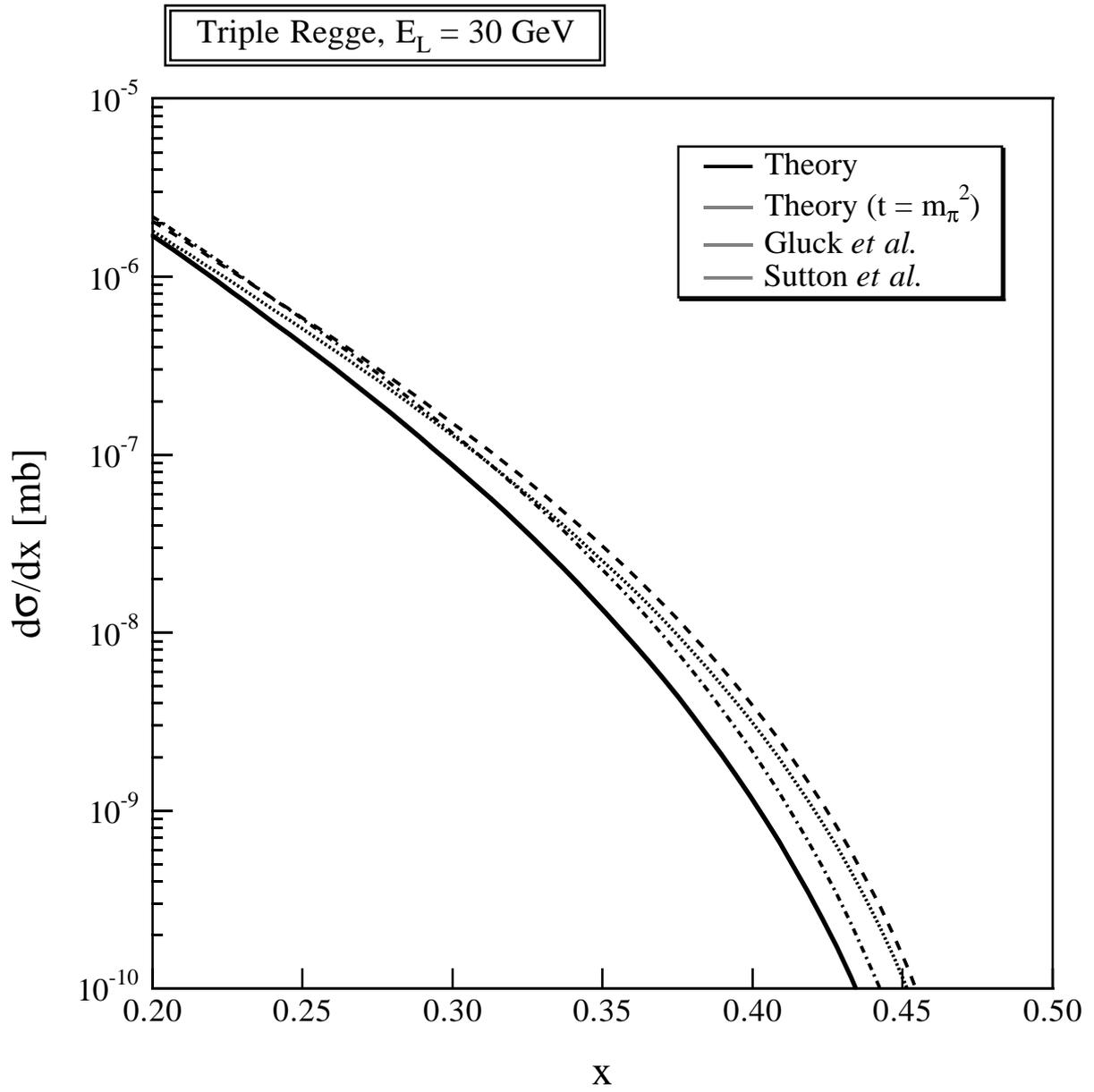
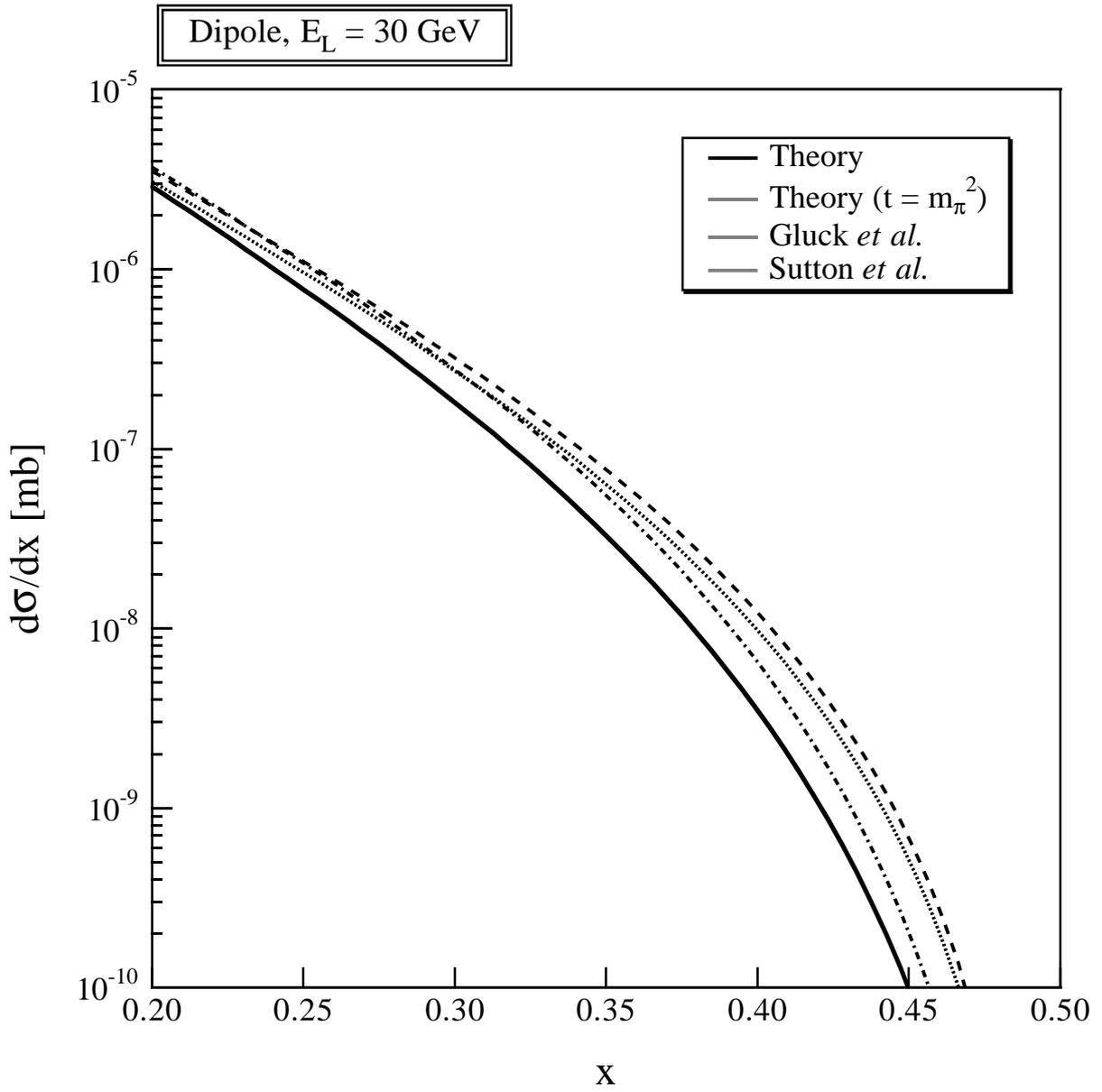
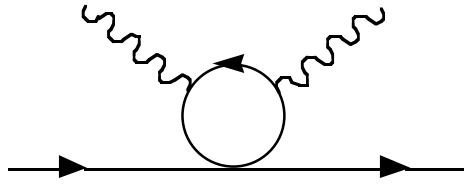
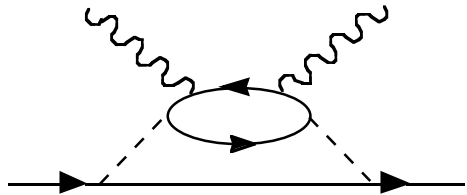
\end{document}